\documentclass[aps,12pt,superscriptaddress,reprint]{revtex4-1}

\usepackage{graphicx,xcolor}

\begin{document}

\title{Ab initio study of point defects in NiTi-based alloys}
\author{David Holec}
\email[]{david.holec@unileoben.ac.at}
\affiliation{Department of Physical Metallurgy and Materials Testing, Montanuniversit\"at Leoben, Austria}
\author{Martin Fri\'ak}
\affiliation{Max-Planck-Institut f\"ur Eisenforschung GmbH, D\"usseldorf, Germany}
\affiliation{Institute of Physics of Materials, Academy of Sciences of the Czech Republic, Brno, Czech Republic}
\author{Anton\'in Dlouh\'y}
\affiliation{Institute of Physics of Materials, Academy of Sciences of the Czech Republic, Brno, Czech Republic}
\author{J\"org Neugebauer}
\affiliation{Max-Planck-Institut f\"ur Eisenforschung GmbH, D\"usseldorf, Germany}

\newcommand{\corr}[1]{{\color{purple}#1}}

\begin{abstract}
Changes in temperature or stress state may induce reversible B2$\leftrightarrow$(R)$\leftrightarrow$ B19' martensitic transformations and associated shape memory effects in close-to-stoichiometric nickel-titanium (NiTi) alloys. Recent experimental studies confirmed a considerable impact of the hydrogen-rich aging atmosphere on the subsequent B2 austenite $\leftrightarrow$ B19' martensite transformation path. In this paper, we employ Density Functional Theory to study properties of Ar, He, and H interstitials in B2 austenite and B19' martensite phases. We show that H interstitials exhibit negative formation energies, while Ar and He interstitials yield positive values. Our theoretical analysis of slightly Ni-rich Ni--Ti alloys with the austenite B2 structure shows that a slight over-stoichiometry towards Ni-rich compositions in a range~51--$52\,\text{at.\%}$ is energetically favorable. The same conclusion holds for H-doped NiTi with the H content up to $\approx6\,\text{at.\%}$. In agreement with experimental data we predict H atoms to have a strong impact on the martensitic phase transformation in NiTi by altering the mutual thermodynamic stability of the high-temperature cubic B2 and the low-temperature monoclinic B19' phase of NiTi. Hydrogen atoms are predicted to form stable interstitial defects. As this is not the case for He and Ar, mixtures of hydrogen and the two inert gases can be used in annealing experiments to control H partial pressure when studying the martensitic transformations in NiTi in various atmospheres.
\end{abstract}

\maketitle

\section{Introduction and motivation}

Near-equiatomic NiTi shape memory alloys (SMAs) belong to the most successful shape memory materials currently used in applications \cite{SMABook}. Their excellent functional and structural properties are based on a good mechanical strength, oxidation and corrosion resistance and reliable shape memory behavior \cite{Saburi1998,Kim1997,Miyazaki1998,Duerig1999,vanHumbeeck1999,Sittner2000}. All the important shape memory effects (the one-way effect, the two-way effect and pseudoelasticity) are primarily associated with the martensitic transformation \cite{Otsuka1998}. The martensitic transformation converts the high temperature parent B2-austenite phase (CsCl-type ordered cubic lattice) into the ``soft'' martensite R-phase ($P3$ trigonal lattice), and further into the low temperature martensite B19'-phase ($P2_1/m$ monoclinic lattice). 

\begin{figure}[b]
  \includegraphics[width=6cm]{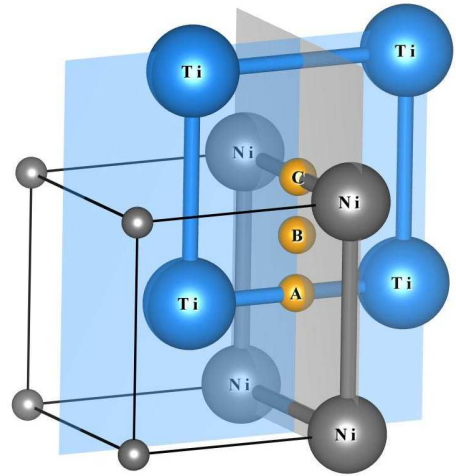}
  \caption{(Color online) Interstitial positions in the cubic B2 NiTi lattice. Larger blue spheres are Ti atoms, smaller gray spheres are Ni atoms. The interstitial positions A, B, and C are marked with the small orange spheres. Blue planes contain only Ti atoms while gray planes are occupied by Ni atoms.}
  \label{fig:int_positions}
\end{figure}

The intermetallic NiTi B2-phase exhibits some solubility for Ni that increases with increasing temperature \cite{BinaryDiagrams}. 
Temperatures of martensitic phase transitions (PTTs) are strongly dependent on the alloy composition and the impurity content \cite{Saburi1998,KhalilAllafi2002,Tang1999,Frenzel2010}. Varying the nickel content in Ni-rich NiTi SMA from 50 to $51\,\text{at.\%}$ allows changing the PTTs by more than $100\,^{\circ}\, \text{C}$ \cite{KhalilAllafi2002,Tang1999,Frenzel2010}.  Therefore, the PTTs can be controlled by the concentration of nickel and thus tailored for a specific application.

\citet{Nishida2003} first provided evidence that variations in heat treatment environments may result in considerable changes of PTTs in subsequent martensitic transformations. They pointed out that the solution annealing is  the critical step as far as the impact of the heat treatment atmosphere on the transformation path is concerned. The path of the martensitic phase transition was shown to change once oxygen was a part of a residual gas during the heat treatment. In order to explain the observed effects, \citet{Fujishima2006} and recently also \citet{Ravari2012} suggested a link between the precipitation of the Ni$_4$Ti$_3$ phase and the composition of the heat treatment atmosphere. While the scenario proposed by \citet{Nishida2003} focused on the role of the residual oxygen, some other gaseous components, that may have contributed to the overall pressure of residual gasses and thus to the varying PTTs, were not taken into account. This is rather surprising since numerous studies showed that other gasses, like e.g. hydrogen, may result in similar effects. 

Many experimental and theoretical studies explored various aspects of the interactions between environments containing hydrogen and NiTi-based intermetallics, see e.g.~\cite{Wayman1964,Biscarini2005,Mazzolai2007,Biscarini1999,Coluzzi2006,Mazzolai2007a}. 
Mazzolai and co-workers investigated and reported important parameters describing solubility and diffusion of interstitial hydrogen in the bulk NiTi SMAs \cite{Biscarini2005,Mazzolai2007}. Furthermore, the same research team applied advanced mechanical spectroscopy techniques \cite{Biscarini1999,Coluzzi2006,Mazzolai2007a,Villa2009} in order to characterize interactions of interstitial hydrogen with dislocations, internal interfaces involved in \corr{martensitic transformations (MTs)}, such as twin boundaries and moving parent-product interface. A  prospective strain-glass hypothesis has also been tested~\cite{Villa2009,Wang2006,Wang2008}. Most importantly, with respect to the objectives of the present study, the internal friction experiments provided a clear evidence that a relaxation peak caused by interactions between hydrogen atoms and the austenite-martensite interface is suppressed for hydrogen concentrations higher than $4.5\,\text{at.\%}$. This experimental result can only be rationalized assuming that higher hydrogen concentrations effectively inhibit the overall B2$\leftrightarrow$R$\leftrightarrow$B19' transformation process. This result received confirmations from both, (i) independent \corr{differential scanning calorimetry (DSC)} experiments \cite{Runciman2008,Kubenova2011,Kubenova2012,Zalesak2013} and (ii) \textit{ab initio} density functional theory (DFT) calculations \cite{Moitra2011}. In passing we note that the decay of B2$\leftrightarrow$R$\leftrightarrow$B19' MT due to the increasing hydrogen content exhibits some similarity to the suppression of the same transformation by increasing dislocation density during thermal transformation cycles \cite{Pelton2012}.

\citet{Moitra2011} used DFT calculations to study bulk and surface diffusion of H in NiTi. In their calculations, octahedral interstitial positions yielded the lowest energy states. \citet{Kang2006} studied NiTiH$_x$  using the \textit{ab initio} tight-binding method and considered also the octahedral interstitial positions with 4 Ti and 2 Ni nearest neighbors in order to form the hydride phases. On the other hand, the MD based study of \citet{Ruda1996} favors H atoms being placed in tetrahedral rather than in octahedral sites. Moreover, it is not clear what impact the local relaxations may have on the site occupation of the H atoms. Hence, the aim of this study is to further explore the stability of point defects, such as H, He, and Ar interstitials or Ni(Ti) or Ti(Ni) anti-sites, in the nearly equiatomic NiTi alloys, and to study their impact on the martensitic transformation process. We employ a state-of-the-art {\it ab initio} modeling procedure based on the Density Functional Theory~\cite{Hohenberg1964,Kohn1965}.

\section{Methodology} \label{sec:int_positions}

The present study focuses mainly on the austenite B2 phase, however, we also discuss briefly the stability of interstitials in the  martensite B19' phase. Table~\ref{tab:lattice} summarizes the structural parameters taken from our previous study~\cite{Holec2011b} that were adopted in the present work.

\begin{table}[t]
  \caption{Lattice parameters used in this study (taken from~\cite{Holec2011b}).}
  \label{tab:lattice}
  \begin{ruledtabular}
    \begin{tabular}{lccccc}
	& $a\,[\mbox{\AA}]$ & $b\,[\mbox{\AA}]$ & $c\,[\mbox{\AA}]$ & $\gamma$ & $x^{\text{Ni}}$ \\
      B2 & $3.007$ & & & $90.0^\circ$ & $1.000$ \\
      B19' & $2.732$ & $4.672$ & $4.234$ & $95.3^\circ$ & $0.980$
    \end{tabular}
  \end{ruledtabular}
\end{table}

In total, there are three non-equivalent interstitial positions in the cubic B2 structure (see also Fig.~\ref{fig:int_positions}). First, the tetrahedral B in which the nearest neighbors are 2 Ni and 2 Ti atoms. Second,  the octahedral position A where the surrounding octahedron is formed by 4 Ni and 2 Ti atoms. Third, the octahedral position C which is characterized by 2 Ni and 4 Ti nearest neighbors. The similar situation exists also in the B19' phase; however, the mutual displacement of atoms with respect to the ideal B2-lattice positions (B19' unit cell contains 2 formula units, while the cubic B2 only one) results in two non-equivalent variants of  the three interstitial positions defined above for the B2 phase.

The calculations are performed using Density Functional Theory (DFT) as implemented in the Vienna Ab initio Simulation Package (VASP) \cite{Kresse1996} together with plane-wave projector augmented wave pseudopotentials \cite{Kresse1999} and the Perdew-Burke-Ernzerhof generalized gradient approximation for the exchange and correlation effects \cite{Perdew1996}. The plane-wave cut-off energy was set to $400\,\textrm{eV}$, while the reciprocal space was sampled with a mesh equivalent to the $24\times24\times24$ $k$-points for the B2 unit cell. Interstitial defects were modeled by inserting a single atom inside $2\times2\times2$ (16 lattice sites), $3\times3\times3$ (54 lattice sites), or $4\times4\times4$ (128 lattice sites) supercells.

The stability of an interstitial atom in the NiTi phase is assessed based on the energy of solution as
\begin{equation}
  E_s=E(\text{NiTi}+X) - \Big[E(\text{NiTi})+E(X)\Big] \label{eq:Es}
\end{equation}
where $E(\text{NiTi}+X)$ and $E(\text{NiTi})$ are the total energies of the NiTi phase with and without the interstitial defect. $E(X)$ is the energy (per atom) of the interstitial in its stable form (\corr{$\text{H}_2$ molecule, He or Ar atom) when put in a large vacuum box with side of $20\,\mbox{\AA}$}. The values of $E_s$ are negative for stable defects, while positive values suggest that the respective element is completely insoluble in NiTi.

\section{Results and discussion}

\subsection{Influence of the local off-stoichiometry}

Before studying the impact of ternary interstitials, we examine thermodynamic properties of both stoichiometric NiTi and slightly off-stoichiometric Ni--Ti alloys. In order to analyze the thermodynamic stability of Ni-Ti alloys with the stoichiometry deviating from that of equiatomic NiTi,we evaluate the energy of formation, $E_f$. It is calculated (per atom) as
\begin{equation}
  E_f=\frac{\Big[E(\text{supercell})-\Big(N_{\text{Ni}}E(\text{Ni}^{\text{fcc}})+N_{\text{Ti}}E(\text{Ti}^{\text{hcp}})\Big)\Big]}{N_{\text{Ni}}+N_{\text{Ti}}}
\end{equation}
where $N_X$ and $E(X)$ are, respectively, the number of atoms $X$ in the alloy  supercell and the energy of an element (per atom) in its stable unary phase. 

\begin{figure}
  \includegraphics{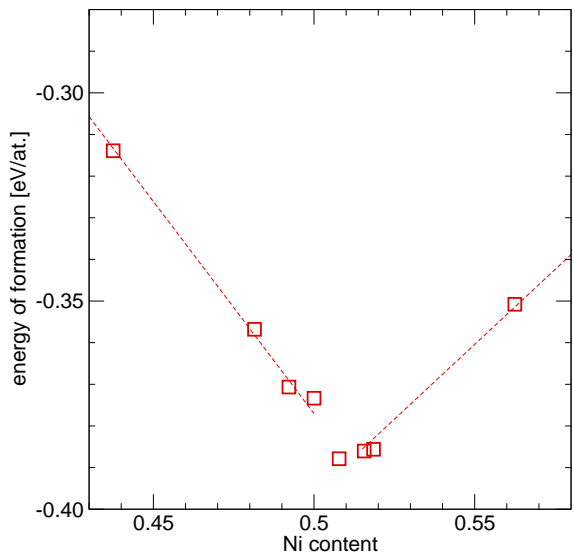}
  \caption{Energy of formation, $E_f$, per atom as a function of Ni content in the Ni$_x$Ti$_{1-x}$ alloy. The dashed lines are linear interpolations through the data for both Ni-rich and Ti-rich alloys around the equiatomic NiTi.}
  \label{fig:Ef}
\end{figure}

The results shown in Fig.~\ref{fig:Ef} demonstrate a tendency of the B2 NiTi phase for a slight off-stoichiometry. The Ni-rich compositions in the range  $51$--$52\,\text{at.\%}$ of nickel exhibit the lowest values of $E_f$. In our model, the off-stoichiometry is realized by anti-site defects Ni(Ti) and Ti(Ni), where some Ti sites are occupied by Ni atoms [labeled as Ni(Ti)], and vice versa. \citet{Lu2007} showed that the Ni(Ti) anti-site defect is energetically more preferred than the Ti-site vacancy. The trend presented in Fig.~\ref{fig:Ef}, which suggests that sparse Ni(Ti) anti-sites are thermodynamically favorable to a perfect stoichiometric crystal, is in agreement with the experimental data, e.g. with the phase diagram in which the NiTi phase field extends towards Ni-rich compositions \cite{BinaryDiagrams}. The effect of composition and off-stoichiometry on the stability of B2 NiTi phase were studied by \citet{Lai2000} using molecular dynamics (MD) simulations. However, the compositional step of $\Delta x=0.1$  used in their work was too large and thus did not allow resolving the formation energy minimum  predicted in the present study for only slightly Ni-rich compositions.

\subsection{Local environments of H interstitials}\label{sec:relaxation}

The three different interstitial positions in the B2 NiTi lattice were described in Section~\ref{sec:int_positions}. The solution energy associated with the presence of H in the Ni-rich octahedral site A of the $2\times2\times2$ supercell is $+0.439\,\text{eV}$ per solute atom, while the solution energy is negative for B ($-0.477\,\text{eV/interstitial}$) and C ($-0.391\,\text{eV/interstitial}$). Consequently  the Ni-rich octahedral site A represents an unstable position while the other two sites B and C can host hydrogen atoms. 

\begin{figure}
  \includegraphics[width=8cm]{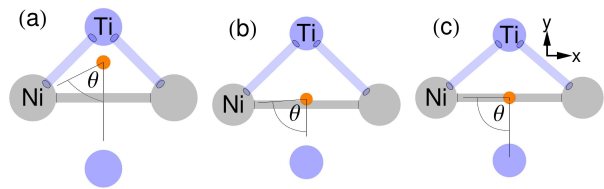}
  \caption{(a) Starting tetrahedral interstitial position with 2 Ni and 2 Ti atoms as the nearest neighbors relaxes into (b) octahedral-like non-symmetrical environment with 2 Ni and 4 Ti atoms. (c) Octahedral interstitial position with 2 Ni and 4 Ti nearest neighbors. The interstitial atom is visualized by the small orange sphere.}
  \label{fig:loc_neighbourhood}
\end{figure}

\begin{figure}
  \includegraphics{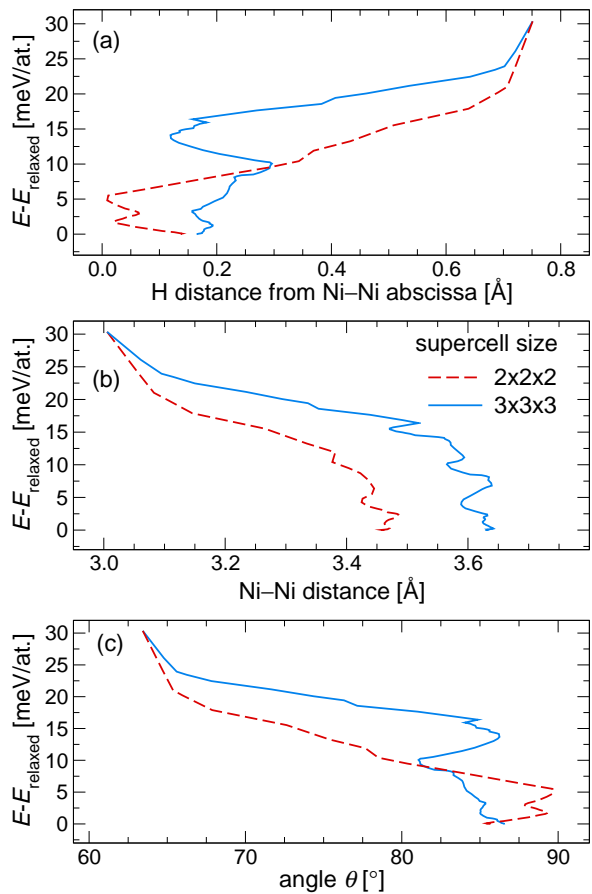}
  \caption{Changes in (a) H atom position, (b) Ni-Ni distance and (c) the angle $\theta$ during the relaxation of H atom from the tetrahedrally coordinated interstitial position B to a Ti-rich octahedral-like relaxed configuration. For further details see the text.}
  \label{fig:relaxation}
\end{figure}

A detailed analysis of the relaxed supercell geometries reveals that the H atom, originally situated in our simulation in the tetrahedral interstitial site B, relaxes towards the octahedral Ti-rich configuration C (see \corr{Fig.~\ref{fig:loc_neighbourhood}}). In order to quantify associated distortions of the surrounding lattice, we analyzed the geometry of the octahedron surrounding the H interstitial. We characterize the geometry by three structural parameters that are related to the solution energy per atom (Eq.~\ref{eq:Es}), i.e. the difference between the total \corr{energies of the current} supercell and the final fully relaxed state, as plotted in Fig.~\ref{fig:relaxation}. These characteristics are: 1)  the distance of the H atom from the Ni--Ni abscissa in the Ti-plane (Fig.~\ref{fig:relaxation}a), 2) the length of the Ni--Ni abscissa (Fig.~\ref{fig:relaxation}b), and 3) the (smallest) angle $\theta$ between the H--Ni direction and the Ti-plane.

Independent of the supercells size (being either $2\times2\times2$ or $3\times3\times3$), the H atom approaches the Ni--Ni abscissa and finds its energetically preferred position $\approx0.15\,\mbox{\AA}$ away from the Ni-Ni line. At the same time, the Ni atoms move further apart and the angle $\theta$ increases from $\approx67^{\circ}$ to $\approx87^{\circ}$ but does not reach 90$^{\circ}$ expected for the highly symmetric interstitial site C. 
We can thus conclude that, even though the tetrahedral site provides an interstitial position with the lowest solution energy (Eq.~\ref{eq:Es}) as compared with other two interstitial sites A and C, the ideal geometry shown in Fig.~\ref{fig:loc_neighbourhood}a (i.e. the starting configuration for structural relaxation) does not represent a stable configuration for the H atom.

\subsection{Energetics of interstitials in B2 phase}

Results of our calculations suggest that even small spatial perturbations are sufficient to start the relaxation process in which H atoms escape from the highly symmetric Ti-rich octahedral sites C and set out to non-symmetrical positions, which, however, resemble the environment of the octahedral Ti-rich site. Energy gains associated width such re-locations from the fully symmetric octahedral Ti-rich C-type atomic position to the non-symmetric one are $86\,\text{meV}$, $457\,\text{meV}$ and $550\,\text{meV}$ for $2\times2\times2$, $3\times3\times3$, and $4\times4\times4$ supercells, respectively. Obviously, the relaxation process towards the non-symmetrical Ti-rich is energetically highly favorable.

\begin{figure}
  \includegraphics{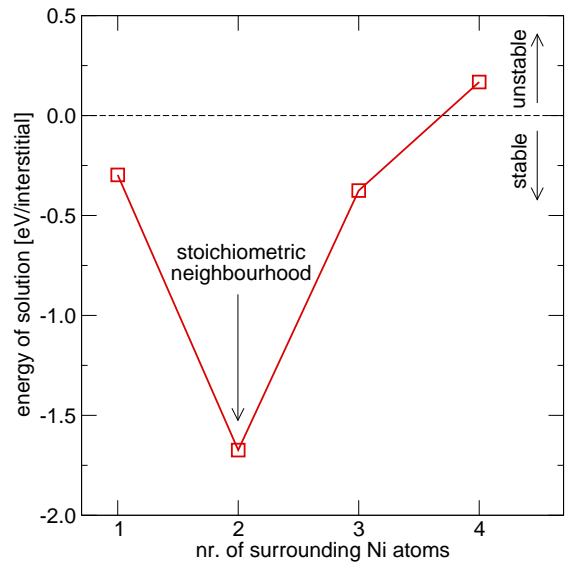}
  \caption{Energy of solution associated with H interstitial in B-type site as a function of its the local environment (anti-site configuration of the parent B2 matrix).}
  \label{fig:Es_offstoichiometry}
\end{figure}

Since the B2 NiTi intermetallics prefers slightly off-stoichiometric Ni-rich compositions realized by Ni(Ti) anti-sites (see Fig.~\ref{fig:Ef} and~\cite{Lu2007}), we have also investigated interactions between H interstitials and Ni(Ti) anti-sites defects. Four starting configurations were considered within the  $4\times4\times4$  supercell in which the H atom in a tetrahedral B-type position was embedded in four different environments. The regular (2Ni,2Ti) tetrahedral geometry (Fig.~\ref{fig:loc_neighbourhood}a) corresponds to the situation when the H interstitial and the \corr{Ni(Ti)} anti-site defects are spatially well separated. In addition, three off-stoichiometric environments (1Ni, 3Ti), (3Ni, 1Ti), and (4Ni, 0Ti) were considered. The corresponding energies of solution (obtained from the $4\times4\times4$ supercells) are plotted in Fig.~\ref{fig:Es_offstoichiometry}. Results presented in Fig.~\ref{fig:Es_offstoichiometry} clearly show that the energetically most favorable H atom environment is a defect-free B2 NiTi matrix rather than a close vicinity of any anti-site defect. In all the four investigated cases, the inspection of the local environments confirms a tendency for relaxation towards non-symmetrical octahedral-like configurations discussed in the preceding section (see Fig.~\ref{fig:loc_neighbourhood}b).

\begin{figure}
  \includegraphics{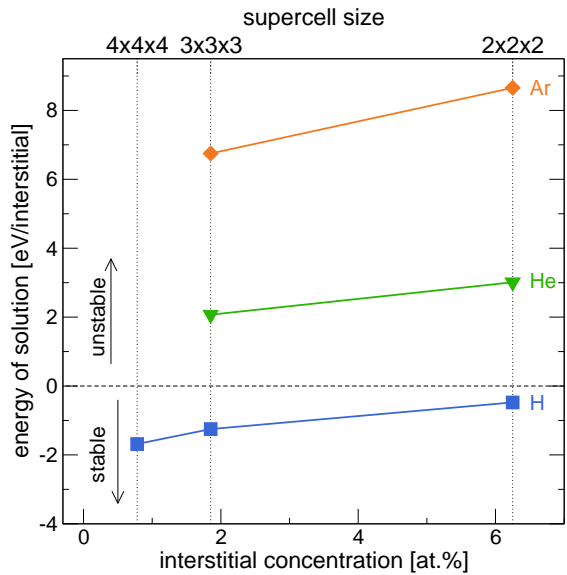}
  \caption{Energy of solution of Ar, He, and H in B2 NiTi phase as a function of the supercell size (interstitial concentration).}
  \label{fig:Es_overview}
\end{figure}

A hydrogen partial pressure required in the annealing experiments 
can be accurately adjusted using mixtures of H with some appropriate inert gas like Ar or He~\cite{Zalesak2013}. It is thus important to assess whether He or Ar atoms are likely to penetrate into the bulk of B2 NiTi phase and form stable interstitials in a way similar to  H atoms.  Corresponding energy of solution calculated according to Eq.~\ref{eq:Es} are plotted in Fig.~\ref{fig:Es_overview}.

The data presented in Fig.~\ref{fig:Es_overview} correspond to the interstitial position with the lowest energy which for all the three elements is the relaxed configuration shown in~Fig.~\ref{fig:loc_neighbourhood}b. The system converges to this relaxed configuration  starting from the interstitial atom being located in the tetrahedral position B  (Fig.~\ref{fig:loc_neighbourhood}a). The energy of solution was evaluated for three sizes  $2\times2\times2$, $3\times3\times3$ and $4\times4\times4$ of the supercell, each time containing one interstitial, thus yielding, respectively, interstitial concentrations of  $\approx6.2$, $\approx1.8$, and $\approx0.8\,\text{at.\%}$ of $X$.

The solution energy $E_s$ per one H interstitial increases from $-1.683$ to $-0.477\,\text{eV per interstitial}$ as the concentration changes from $\approx0.8$ to $\approx6.2\,\text{at.\%}$. The fact that the values do not settle even for the lowest $H$ concentration means that even with the largest supercell containing 129 atoms we still have not  reached the dilute limit for the interstitial energy. The energy drop with the increasing supercell size is related to an easier and more efficient absorption of the elastic distortions caused by the interstitial.

When both, the internal coordinates of atoms and a supercell shape are allowed to optimize during the calculation, the total energy  further decreases by $\approx100$--$200\,\text{meV}$ for the $2\times2\times2$ supercells, while only small negative gain up to $10\,\text{meV}$ is obtained in the case of the $3\times3\times3$ supercell. This results suggests that an accurate model of B2 NiTi with $\approx6\,\text{at.\%}$ of interstitial H should consider larger than $2\times2\times2$ supercell with more than one H atom.

In contrast to rather reactive H atoms, chemically inert Ar and He atoms are predicted to be soluble in NiTi at \corr{neither $1.8\,\text{at.\%}$ nor $\approx6\,\text{at.\%}$ concentration}. We suspect that the higher energy of solution in case of Ar (and therefore its much lower solubility) is mostly due to the larger atomic size of Ar ($r_{\text{He}}=128\,\text{pm}$, $r_{\text{Ar}}=174\,\text{pm}$).
Importantly for experimental studies, neither Ar nor He is expected to be found in NiTi after annealing and these gases can thus be employed to fine tune a required partial pressure of hydrogen.

\subsection{Influence of H on the B2-B19' energy difference}

\begin{figure}
  \includegraphics{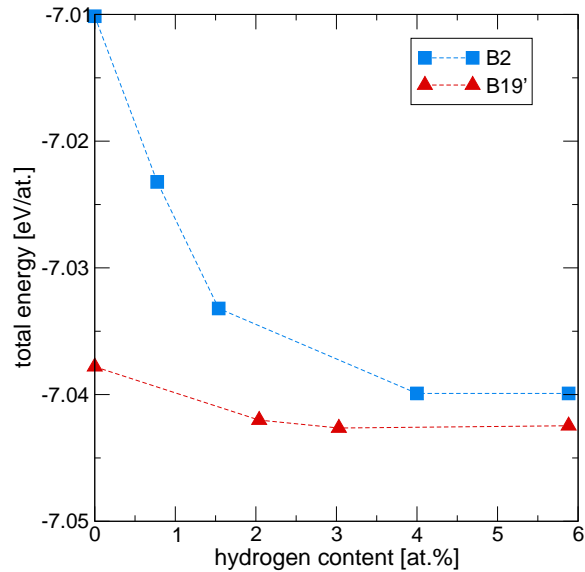}
  \caption{The total energies of the martensite B19' and austenite B2 phases as a function of the interstitial H content.}
  \label{fig:Ediff}
\end{figure}

The strong impact of H on the B2$\leftrightarrow$B19' transformation path has been confirmed experimentally~\cite{Runciman2008,Kubenova2011,Kubenova2012,Zalesak2013}.
A question thus rises  whether the \textit{ab initio} simulations can contribute to a better understanding of this phenomenon. Since the experimentally observed suppression of B2$\leftrightarrow$B19' transformation in hydrogen charged NiTi alloys may be associated with changes in the thermodynamic stability of the corresponding austenite and martensite lattices, we have calculated the total energies of B2$\leftrightarrow$B19' phases with containing different content amount of H interstitials.

The results plotted in Fig.~\ref{fig:Ediff} demonstrate that the total energies of the two phases come closer with increasing H content. This trend may cause a drop of the driving force for the transformation and may thus extend the B2 stability range into the low temperature regime. These data points presented in Fig.~\ref{fig:Ediff} further suggest that a critical H concentrations may exist at which the two energy curves cross each other and the B2 phase may exhibit better stability than the B19' phase for H concentrations beyond the critical point.  These results are fully in line with the experimental observations showing that the martensitic transformation is ceased for interstitial H content higher than about $4.5\,\text{at.\%}$~ \cite{Coluzzi2006,Mazzolai2007a,Mazzolai2007a,Villa2009,Runciman2008,Kubenova2011,Kubenova2012,Zalesak2013}.

In this context it should be noted that the behavior of H interstitials in the B2 and B19' is qualitatively very similar: H atoms starting in a B19' tetrahedrally coordinated positions relax into octahedral-like environments with 6 nearest neighbors. Furthermore, the lowest energy configurations, which yield the B19' data shown in Fig.~\ref{fig:Ediff},  are always Ti-rich, i.e. with 4 nearest neighbors being Ti atoms. 
	
\begin{table*}
  \caption{Inclination $\theta$ of the H--Ni directions from the plane containing the Ti atoms in the octahedrons around the H interstitial atoms.}
  \begin{ruledtabular}
  \begin{tabular}{lllllll}
      & ideal B2 & tetrahedral B2 & octahedral B2 & shuffled octahedral B2 & ideal B19' & tetrahedral B19' \\
    $\theta$ & $90.0^\circ$ & $87.3^\circ$ & $90.0^\circ$ & $86.0^\circ$ & $81.7^\circ$ & $82.5^\circ$\\
  \end{tabular}
  \end{ruledtabular}
  \label{tab:inclinations}
\end{table*}

The structural similarity between B2 and B19' phases induced by H interstitials, can also be documented using the inclination angle $\theta$ (introduced in Section~\ref{sec:relaxation}) between the H--Ni direction and the plane containing Ti atoms. The results angles $\theta$  calculated for several configurations are summarized in Table~\ref{tab:inclinations}. 
As compared to the fully relaxed hydrogen-free B2 and B19' lattices, the relaxation of hydrogen from the B2 tetrahedral B-sites to the non-symmetrical sites (Fig. 3b) tends to decrease the inclination from 90$^{\circ}$ down to 87.3$^{\circ}$ while the opposite trend is observed for the B19' phase where the inclination grows from 81.7$^{\circ}$ to 82.5$^{\circ}$. 
Consequently, the H interstitial locally smears out (to a certain extent) the structural differences between the austenite B2 and martensite B19' phases. A similar effect is also obtained for another starting configuration where the H atom is situated in the octahedral position C in the B2 phase. Here, however, the initial positions of the 4 Ti and 2 Ni atoms are slightly shifted along the $[011]$ directions before the structural relaxation takes place. We refer to this case as the shuffled C octahedral configuration.

The structural relaxation of the H atom in the shuffled octahedral interstitial site yields energy and geometry comparable with the case when starting configuration is represented by the tetrahedral coordination. However, unlike in the case of the non-symmetrical site (see Fig.~\ref{fig:loc_neighbourhood}b), the two Ni and one H atoms fall on a straight line for the shuffled octahedral geometry. Thus, the shuffled C octahedral position resembles the B19' structure even better than the relaxed tetrahedral B positions. Similar mechanism has been recently explored by \citet{Tahara2011} in oxygen doped $\beta$-Ti alloys.

\section{Conclusions}

Quantum-mechanical calculations were performed to study interstitials and anti-site point defects in NiTi phases. Using the \textit{ab initio} techniques, we show that a slight Ni-rich off-stoichiometry of the B2 NiTi phase is energetically favorable when it is realized by the Ni(Ti) anti-site defects. Regarding interstitials solutes, our results clearly show that H atoms have a strong impact on the martensitic phase transformation in NiTi by altering the mutual thermodynamic stability of the high-temperature cubic and the low-temperature low-symmetry phase of NiTi. Hydrogen atoms are predicted to form stable interstitial defects, unlike He and Ar atoms. This opens a possibility to use mixtures of hydrogen and the two inert gases in annealing experiments in which effects of controlled H partial pressure on the subsequent martensitic transformations are investigated. Hydrogen atoms preferably occupy non-symmetrical interstitial positions, similar to the octahedral site C with 4 Ti and 2 Ni nearest neighbors. Characteristics of the local lattice distortions around the H interstitial defects in the B2 and B19' phases as well as a decreasing difference in the corresponding B2 and B19' total energies are in line with the extended stability of the B2 phase observed in the martensitic transformation experiments. In conclusion, our study provides insights into some fundamental processes governing martensitic transformation in NiTi exposed to realistic conditions such as gaseous atmospheres.

\section{Acknowledgments}

A.D. acknowledges funding by the CSF, project no. 106/09/1913, and an additional financial support through the IPM AS CR development program no. RVO:68081723. M.F. acknowledges finacial support from the Academy of Sciences of the Czech Republic through the Fellowship of Jan Evangelista Purkyn{\v e}.

%

\end{document}